\documentclass[useAMS,usenatbib]{mnras}
\usepackage{graphicx}\usepackage{epsfig}\usepackage{epsf}
\usepackage{amsmath}
\usepackage{amssymb}
\usepackage{stfloats}
\usepackage{orcidlink}
\usepackage{lipsum}
\usepackage{pgf}
\usepackage{tikz} 
\usetikzlibrary{shapes,arrows,automata}
\usepackage{scrextend}
\usepackage{tree-dvips}
\usepackage{soul}

\def\hst{{\it HST}}

\title[SN~2009ip's progenitor is gone]{SN 2009ip after a decade: The luminous blue variable progenitor is now gone}

\author[Smith et al.]{Nathan Smith \orcidlink{0000-0001-5510-2424},$^1$\thanks{E-mail: nathans@as.arizona.edu} 
Jennifer E.\ Andrews \orcidlink{0000-0003-0123-0062},$^{2}$ 
Alexei V.\ Filippenko \orcidlink{0000-0003-3460-0103},$^{3}$
Ori~D.~Fox \orcidlink{0000-0003-2238-1572},$^{4}$
\newauthor Jon C. Mauerhan \orcidlink{0000-0002-7555-8741},$^{5}$
and Schuyler D.~Van Dyk \orcidlink{0000-0001-9038-9950}$^{6}$ \\
$^1$Steward Observatory, University of Arizona, 933 N. Cherry Ave., Tucson, AZ 85721, USA \\
$^2$Gemini Observatory, 670 N. Aohoku Place, Hilo, Hawaii, 96720, USA \\
$^3$Deparment of Astronomy, University of California, Berkeley, CA 94720-3411, USA \\
$^4$Space Telescope Science Institute, 3700 San Martin Dr., Baltimore, MD 21218, USA \\
$^5$Physical Sciences Laboratory, The Aerospace Corporation M2-266, PO Box 92957 Los Angeles, CA 90009, USA \\
$^6$Caltech/IPAC, Mailcode 100-22, Pasadena, CA 91125, USA
}

\begin{document}
\pagerange{\pageref{firstpage}--\pageref{lastpage}} \pubyear{2021}
\maketitle
\label{firstpage}

\begin{abstract}
    We present new {\it Hubble Space Telescope} (\hst) imaging photometry for the site of the Type IIn supernova (SN) 2009ip taken almost a decade after explosion.  The optical source has continued to fade steadily since the SN-like event in 2012.  In the F606W filter, which was also used to detect its luminous blue variable (LBV) progenitor 13~yr before the SN, the source at the position of SN~2009ip is now 1.2~mag fainter than that quiescent progenitor.  It is 6--7~mag fainter than the pre-SN outbursts in 2009--2011. This definitively rules out a prediction that the source would return to its previous state after surviving the 2012 event.  Instead, the late-time fading matches expectations for a terminal explosion.  The source fades at a similar rate in all visual-wavelength filters without significant colour changes, therefore also ruling out the hypothesis of a luminous dust-obscured survivor or transition to a hotter post-LBV survivor. The late-time continuum with steady colour and strong H$\alpha$ emission detected in a narrow F657N filter are, however, entirely expected for ongoing shock interaction with circumstellar material in a decade-old core-collapse SN.  Interestingly, the ultraviolet flux has stayed nearly constant since 2015, supporting previous conjectures that the F275W light traces main-sequence OB stars in an underlying young star cluster.  We expect that the visual-wavelength continuum will eventually level off, tracing this cluster light.   Without any additional outbursts, it seems prudent to consider the 2012 event as a terminal SN explosion, and we discuss plausible scenarios.
\end{abstract}


\section{INTRODUCTION}\label{sec:intro}

Interacting supernovae (SNe), most commonly classified as Types~IIn and Ibn, explode in dense circumstellar cocoons of their own making, producing bright, narrow emission lines when circumstellar material (CSM) is hit by the shock wave (see \citealt{smith17review} for a review).  Most normal SNe expand and quickly cool, radiating only $\sim$1\% of their explosion energy \citep{arnett96,pp15}.  In SNe~IIn and Ibn, however, strong CSM interaction can efficiently convert ejecta kinetic energy into heat and radiation, powering diverse displays that include superluminous SNe \citep{fa77,sm07}, and SNe that continue shining far longer than radioactive decay would allow \citep{cd94,stritz12,andrews17,fox13,fox20,koss14,smith17}.  From this CSM interaction luminosity, one can infer the pre-SN mass-loss rate.   SNe~IIn generally require extreme pre-SN mass loss, far beyond what normal massive-star winds can supply, evoking comparisons to eruptive luminous blue variables (LBVs) or the most extreme red supergiants (see \citealt{smith14araa} for a review).  However, the most direct link between SNe~IIn and LBVs was provided by the remarkable case of SN~2009ip.

Although SN~2009ip was discovered after brightening in 2009, that event was not a terminal SN explosion, but it turned out to be part of a remarkable sequence of pre-SN activity.   Archival {\it Hubble Space Telescope} (\hst) images in 1999 revealed a very luminous progenitor star ($M_V \approx -9.8$~mag), inferred to have an initial mass of 50--80~$M_{\odot}$ \citep{smith10}.  Even with no bolometric correction, this is more luminous than the observed upper luminosity limit for red supergiants at log($L$/L$_{\odot}$) = 5.5 or $M_{\rm Bol}$ = $-$9~mag \citep{dcb18,mcdonald22}, suggesting that it was an LBV or related blue supergiant \citep{smith10}.

About 5~yr later, it began a slow brightening by $\sim$1~mag and subsequently faded over the next few years, reminiscent of the S~Dor eruptions typically experienced by LBVs \citep{smith10}.  In 2009, upon its discovery as a new transient, it brightened rapidly to $M_V \approx -15$~mag and faded again \citep{smith10}, evoking comparisons to the luminosity spikes preceding the 19th century Great Eruption of $\eta$~Carinae that occurred at times of grazing periastron passes in that eccentric binary \citep{S11,sf11,sk13}.  The spectra during the pre-SN eruptions of SN~2009ip showed strong, narrow (550~km~s$^{-1}$) emission lines, although a few lines showed weak, broad, blueshifted absorption wings out to high velocities of $\sim$10,000~km~s$^{-1}$ \citep{smith10,pastorello13}.  Once again, this is similar to some LBVs like $\eta$~Car, where the bulk of the eruptive outflow is $\sim$600~km~s$^{-1}$ or less, but a small fraction ($\sim$1\%) of material is accelerated to high speeds of 5000--10,000~km~s$^{-1}$ \citep{smith18fast,smith08}.  The slow outflow speed and peak luminosity were comparable to those of previous examples of ``SN impostors" \citep{svd07,svd12,smith11lbv}.

Then, in 2012, SN~2009ip experienced a sea change. In 2012 July to September, it brightened again to an absolute magnitude around $-$15, but this time it was a more sustained brightening.  More importantly, the optical spectrum's character changed substantially, now showing the classic H and He broad P~Cygni absorption and emission profiles that are characteristic of ejecta photospheres in genuine core-collapse SNe, with expansion speeds of $\sim$13,000~km~s$^{-1}$ \citep{mauerhan13}.  The spectrum closely resembled that of a normal SN~II-P, except with narrow CSM emission lines superposed atop the broad-line profiles.   These properties announced that the SN impostor had transitioned into a real SN \citep{mauerhan13,sm12}.  After a brief decline in brightness in late September, SN~2009ip suddenly brightened in a few days to become brighter than typical core-collapse SNe \citep{prieto13}.  

During the object's $-$18~mag peak in October 2012, its spectrum looked like that of a normal SN~IIn, although broad lines from fast SN ejecta reappeared in spectra during the decline from peak luminosity.  The initial fainter brightening in August/September is usually referred to as the ``2012a'' event, and the main luminosity peak in October as the ``2012b'' event.  \citet{mauerhan13} proposed a scenario wherein the 2012a event was a SN explosion of an LBV, which was initially fainter than typical SNe~II-P because of the blue progenitor's smaller radius as compared to red supergiants, and then the bright 2012b event marked the onset of very strong CSM interaction when fast ejecta from 2012a caught up to CSM produced by the preceding LBV-like eruptions.  \citet{mauerhan13} noted that the delay of $\sim$40~days between the 2012a and 2012b peaks made sense, because the 2012a ejecta were moving $\sim$10 times faster than the material ejected about a year earlier in slower LBV-like eruptions --- i.e., they should overtake the previously ejected material in $\sim$0.1~yr. In this scenario, the explosion date would be at the start of the 2012a event.

This interpretation of the 2012 event as a SN was supported by a variety of observational evidence discussed in subsequent studies.  \citet{smith14} pointed out that while a lower-energy explosion within dense CSM could, in principle, power the observed light curve, it could not provide a self-consistent explanation for the observed behaviour including the spectra.  \citet{smith14} established that a SN-like explosion with roughly $10^{51}$~erg was needed to explain the persistence of the very broad P~Cygni lines in the spectrum, and that these were consistent with a normal SN~II photosphere formed in fast ejecta.  At all epochs during the 2012a and 2012b events, the observed spectrum of SN~2009ip was consistent with a combination of narrow CSM lines added on top of the spectral evolution of SN~1987A at similar epochs, and was unlike lower-energy SN impostors.  Moreover, \citet{smith14} pointed out that fast ejecta still seen in the spectrum after the main 2012b peak required that much of the fast ejecta did not participate in CSM interaction (i.e., they were not decelerated), thus requiring significant asymmetry and again implying a $10^{51}$~erg  energy budget.  The 2012a+2012b light curve of SN~2009ip was consistent with published models of core-collapse explosions of blue supergiants developed for SN~1987A, except  with additional CSM interaction luminosity during 2012b \citep{smith14}.  Like SN~1987A, the initial 2012a peak of SN~2009ip was likely fainter than typical SNe~II-P from red supergiants because the blue progenitor had a relatively small radius \citep{smith14,mauerhan13}.  On the other hand, the fast expansion speeds of 13,000 km~s$^{-1}$ seen in SN~2009ip directly contradict expectations for the often suggested non-terminal model: a pulsational pair instability eruption, for which bulk expansion speeds of only 2,000 km s$^{-1}$ are predicted from a variety of models \citep{woosley17}.  The 2012a peak might also have been intrinsically more luminous than $-15$~mag because dusty CSM surrounded the progenitor; an infrared (IR) excess was seen during the 2012a event that faded in the 2012b event \citep{smith13}.  

Spectropolarimetry provided definitive evidence that a $10^{51}$~erg explosion was needed to power SN~2009ip; the highly asymmetric CSM geometry inferred from polarization meant that only a small fraction of the SN ejecta would interact with the disc-like CSM, thus elevating the energy budget compared to what is needed to power the light curve alone \citep{mauerhan14}.  Further study of the spectropolarimetric evolution confirmed this high degree of asymmetry in the disc-like CSM \citep{reilly17}.   Other studies of the photometric and spectral evolution provided a variety of additional evidence that the 2012 explosion of SN~2009ip was consistent with a core-collapse SN~IIn event. For example, \citet{graham14,graham17} found that SN~2009ip resembled many other SNe~IIn at late times, with broader lines than any LBV or SN impostor, and its broad emission lines of Ca~{\sc ii} resembled those in core-collapse SNe.  Late-time spectra of SN~2009ip also showed coronal lines \citep{fox15} like those seen in energetic core-collapse SNe such as SN~2005ip \citep{smith09sn05ip} and SN~1995N \citep{fransson02}, but not seen in SN impostors.  From the 2012b peak onward, SN~2009ip resembled normal SNe~IIn; what was most remarkable about it was its detailed record of pre-SN photometry \citep{mauerhan13,smith14}.   Such data are not available for most events, and \citet{bilinski15} showed that SN~2009ip-like precursor eruptions would have been easily missed in most previous SN searches.   Altogether, the energy budget and other clues pointed toward a core-collapse SN or other terminal event, since it is difficult to understand how any star could survive a $10^{51}$~erg explosion.

If one considers only the radiated energy seen in the light curve --- and one ignores the observed high velocities and their persistence, other clues in the spectral evolution, and the high degree of asymmetry indicated by spectral evolution (line profiles) and polarization --- then it might seem that SN~2009ip could arise from a relatively low-energy explosion of only $10^{50}$~erg or less.  CSM interaction is, after all, an efficient way to convert kinetic energy into radiation.  This led several authors to propose that the 2012b event, which had an integrated radiated energy of $\sim$3$\times$10$^{49}$~erg, was not a true core-collapse SN and that the progenitor would survive the nonterminal event \citep{pastorello13,fraser13,margutti14}.  \citet{moriya15} concluded that the 2012b event was not a core-collapse SN based on a CSM interaction model for a few $\times 10^{49}$~erg explosion that could fit the late-time ($>$200~d) light curve, although this model did not fit the 2012a or 2012b peaks, and it assumed spherical symmetry and velocities that were inconsistent with observations. Motivated by these claims of a $10^{50}$~erg (or less) event and a comparison to much less luminous merger events like V838 Mon and V1309 Sco, a merger model that may include accretion luminosity and jet-driven CSM interaction for SN~2009ip has also been proposed \citep{sk13,kashi13,ts13}.   \citet{fraser13} cited an inferred upper limit to the $^{56}$Ni mass of 0.02~$M_{\odot}$ for SN~2009ip as evidence against a core-collapse event.  However, this depends on the assumed time of explosion: \citet{fraser13} assumed that the explosion was at the start of the 2012b event, but if the explosion was at the beginning of 2012a as proposed by \citet{mauerhan13}, then a $^{56}$Ni mass somewhat more than 0.04~$M_{\odot}$ is allowed \citep{smith14}.  In any case, such values of the $^{56}$Ni mass do not disfavor core collapse, since the median $^{56}$Ni mass inferred for normal SNe~II is only 0.03~$M_{\odot}$ \citep{anderson19}.   Similarly, while the total radiated energy of 3$\times$10$^{49}$~erg is less than $10^{51}$~erg, this is hardly an argument against a core-collapse event, since the total radiated energy is only a lower limit to the energy of the explosion.  Normal core-collapse SNe only radiate $\sim$10$^{49}$~erg, and the total radiated energy for SN~1987A was only $\sim$8$\times$10$^{48}$~erg \citep{mm88}.  Various other observational issues were raised, mainly highlighting how SN~2009ip did not conform to expected norms for typical core-collapse SNe~II-P without strong CSM interaction. There is, of course, no established precedent for what core-collapse SNe from such massive stars {\it should} look like.   Without a detected neutrino burst, it was difficult to prove definitively that SN~2009ip was a core-collapse SN and not some other sort of energetic explosive event.  This lingering ambiguity is largely a reflection of the frustrating fact that CSM interaction can effectively mask the traditional core-collapse signatures that observers are accustomed to seeing.  

Debate about the nature of the 2012 event has persisted.\footnote{Some other transients closely resemble SN~2009ip.  For example, \citet{smith13} pointed out that the double-peaked light curve of SN~2010mc \citep[see also][]{ofek13} was nearly identical to the 2012a+2012b light curve of SN~2009ip.  SN~2015bh was another very similar transient that also appeared to have an eruptive LBV-like progenitor \citep{eliasrosa16,thone17,bg18}.  Gaia16cfr was also similar \citep{kilpatrick18}. Since SN~2009ip is the prototype for this subclass of SNe~IIn, the question about the ultimate fate of the progenitor of SN~2009ip may impact their interpretation as well.}  Despite vociferous claims on either side of the ``terminal-or-not" argument, most researchers have agreed that the late-time behaviour should provide a definitive test.\footnote{It has not escaped us that this expectation is not really a definitive test of the explosion mechanism, and is not based on any physical model for how a massive star actually recovers from a low-energy explosion.  It does, nevertheless, provide a simple testable prediction.}  If SN~2009ip returns to its progenitor's luminosity or continues its eruptive LBV-like variability, then the star probably survived in some form, whereas if it just continues to fade below the progenitor luminosity, then it is likely dead, as in the case of many core-collapse SN progenitors that have since disappeared.\footnote{Several detected progenitors of SNe II-P have faded after the SN \citep{smartt15}.  Among more massive detected progenitors of SNe~IIn, only two besides SN~2009ip have faded.  These are SN~1961V \citep{smith11lbv,kochanek11} and SN~2005gl \citep{gl09}, although SN~1961V may have been a pulsational pair instability SN \citep{ws22}.}
\citet{fraserAtel} announced that SN~2009ip had returned to its progenitor's brightness level, apparently confirming that SN~2009ip had survived the 2012 event.  Time proved otherwise, however, because subsequent observations have shown that SN~2009ip has continued its uninterrupted steady fading up to the present epoch.

In this paper, we present new late-time observations of SN~2009ip, finding that it has now definitively faded well beyond the brightness level of its quiescent progenitor seen in 1999, before the pre-SN eruptive variability began.  Moreover, it has faded at a steady rate, with no sign of continued LBV-like variability.   In Section~\ref{sec:obs} we present new multifilter photometry for the point source at the location of SN~2009ip from late-time \hst\, images. Section~\ref{sec:results} presents the light curve and late-time spectral energy distribution (SED), and in Section~\ref{sec:conclusion} we discuss implications for the nature of the explosive event in 2012 and the demise of its progenitor star.

\begin{table}
\begin{center}\begin{minipage}{2.6in}
    \caption{New \hst\ 
    Images of SN~2009ip}\scriptsize
    \begin{tabular}{@{}lllc}\hline\hline
Date          &Dataset ID      &Filter &Exp. (s) \\  \hline
2021 Dec. 14.1 &IEQE01010/11    &F275W  &4960    \\
2021 Dec. 14.4 &IEQE03021       &F555W  &1034    \\
2021 Dec. 14.4 &IEQE03031       &F606W  &634     \\
2021 Dec. 14.2 &IEQE02010/11/12 &F657N  &10688   \\
2021 Dec. 14.4 &IEQE03011       &F814W  &634     \\
\hline
\end{tabular}\label{tab:obs}
\end{minipage}\end{center}
\end{table}

\begin{table*}
\begin{center}\begin{minipage}{4.7in}
    \caption{Progenitor and Late-time \hst\ photometry of SN~2009ip, including photometry from archival images and new images presented here. Values listed are Vega magnitudes.}\scriptsize
    \begin{tabular}{@{}lccccccccccc}\hline\hline
Date       &Phase  &F275W  &1$\sigma$ &F555W &1$\sigma$ &F606W &1$\sigma$ &F657N &1$\sigma$ &F814W &1$\sigma$ \\  \hline
1999 Jun 29    &$-$13 yr &...    &...   &...    &...   &22.09  &0.01   &...    &...   &...    &...   \\
2015 May 23/25 &1023 d &21.82  &0.02  &22.49  &0.01  &...    &...   &18.87  &0.01  &22.26  &0.01  \\
2016 Sep 14    &1495 d &...    &...   &...    &...   &21.95  &0.01  &...    &...   &22.36  &0.01  \\
2019 Jul 31    &2551 d  &...    &...   &23.34  &0.02  &...    &...   &...    &...   &22.97  &0.02  \\
2021 Dec 14    &3417 d  &21.92 &0.02  &23.73  &0.02  &23.26  &0.02  &21.08  &0.01  &23.47  &0.04  \\
\hline
\end{tabular}\label{tab:phot}
\end{minipage}\end{center}
\end{table*}

\section{OBSERVATIONS}\label{sec:obs}

\begin{figure}
\begin{center}
\includegraphics[width=3.3in]{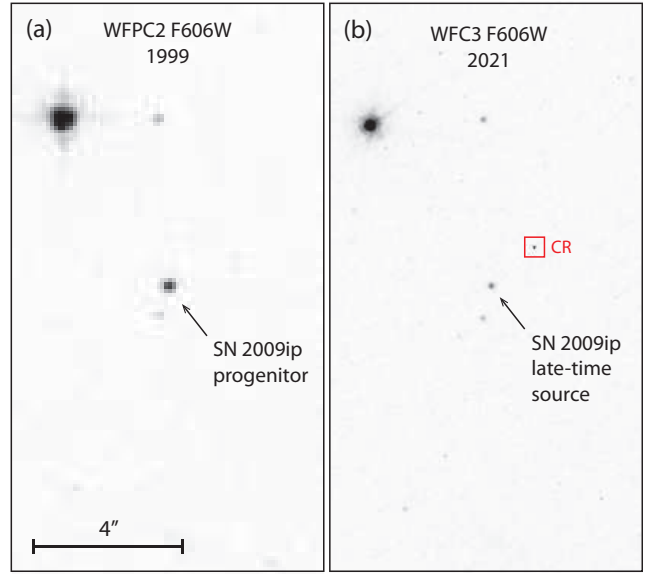}
\caption{{\it HST} images in the same F606W filter showing (a) the 1999 progenitor from \citet{smith10}, as well as (b) the 2021 December image of the late-time fading source at the position of SN~2009ip.  Note that the 1999 image is taken with the WF chip of the WFPC2 camera, and has a larger pixel scale than the dithered and drizzled 2021 WFC3 image.  The source in the red box in the 2021 image is a cosmic ray residual.}\label{fig:img}
\end{center}
\end{figure}

We observed the site of SN~2009ip using \hst/WFC3-UVIS in December 2021 during Cycle 29 (PI Smith, GO-16649).  Our goal was to measure the continued rate of late-time fading of SN~2009ip.   We aimed to duplicate earlier deep multifilter observations taken in 2015 as part of our program to study the environment around SN 2009ip (PI Smith, GO-13787) using four filters: F275W, F555W, F657N (H$\alpha$), and F814W.  For further details about those 2015 observations and an analysis of the environment around SN~2009ip, see \citet{smith16sn09ip}.   We also observed SN~2009ip in the F606W filter, because it was used for the WFPC2 images taken in 1999 that detected the quiescent progenitor star discovered by \citet{smith10}.   The new \hst\, imaging observations obtained in December 2021 are summarised in Table~\ref{tab:obs}.  From the \hst\, archive we also obtained and analyzed unpublished ACS/WFC F606W and F814W images obtained in 2016 (PI Fraser, GO-14150) and WFC3/UVIS F555W and F814W images obtained in 2019 (PI Filippenko, GO-15166).  Figure~\ref{fig:img} shows the new F606W image compared to the 1999 image of the progenitor in that same filter from \citet{smith10}.  The other images look similar to the 2015 images taken in the same F275W, F555W, F657N, and F814W filters, for which representative images were already shown in Figure~1 of \citet{smith16sn09ip}.

\begin{figure*}
\begin{center}
\includegraphics[width=7.0in]{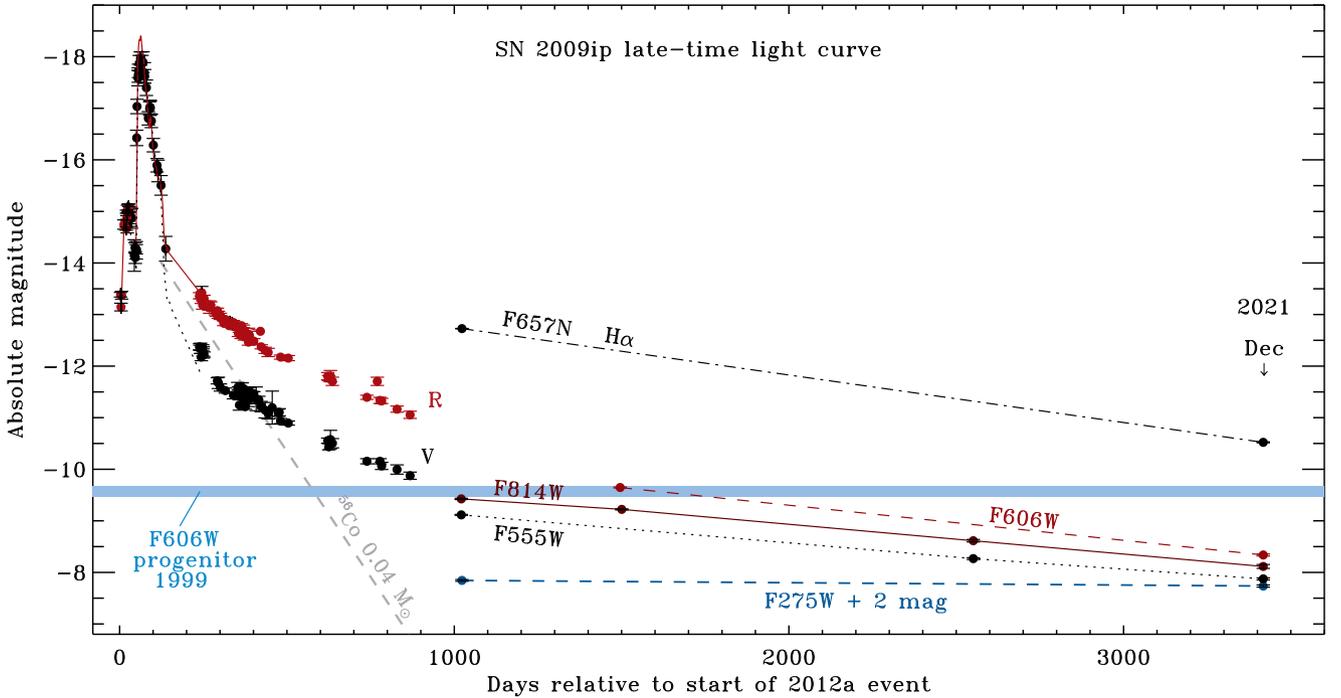}
\caption{The late-time light curve of SN~2009ip, including ground-based
photometry during the main SN-like peak of the 2012 event and shortly afterward, and \hst\, photometry after day 1000; this is similar to Figure 1 of \citet{smith14}, but extended in time to the present epoch.   The ground-based photometry is shown with the $V$ band in black (solid points and dashed black line) and the $R$ band in red (only a solid red line at early times, for clarity).  The $V$ and $R$ magnitudes during the main peak and afterward are from \citet{margutti14} and \citet{smith14} (see \citealt{smith14} for additional details).   The $V$ and $R$ magnitudes for days 200--1000 are from \citet{fraser15} and \citet{smith14}. The late-time \hst\, photometry is shown for the F275W (blue, dashed), F555W (black, dotted), F606W (red, dashed), F657N (black, dot-dashed), and F814W (red-brown, solid) filters; these are from our Table~\ref{tab:phot}.  F275W is shifted by +2~mag for clarity.  The horizontal blue bar shows the F606W magnitude of the quiescent progenitor star from \citet{smith10}.}\label{fig:lc}
\end{center}
\end{figure*}

Table~\ref{tab:phot} lists the resulting Vega-based magnitudes for the point source at the position of SN~2009ip measured in these \hst\, images. The photometry was done for these data using \textsc{dolphot}\footnote{http://americano.dolphinsim.com/dolphot/} \citep{dolphot02,dolphin2016} on the FLC frames (*.flc files are standard \textsc{stsci} preprocessed UVIS-calibrated and CTE-corrected
frames). We used the recommended parameters for \textsc{dolphot} as adopted from \citet{Dalcanton12}, including values for the parameters \texttt{FitSky}=3 and \texttt{RAper}=8 for the photometry.  Note that we remeasured the apparent magnitude of the 1999 progenitor using the same methods as for the later data so that the analyses of the progenitor and late-time source are as consistent as possible.  The F606W magnitude we find for the 1999 progenitor (22.09$\pm$0.01) is marginally fainter than the value (21.8$\pm$0.2) reported by \citet{smith10}, but this is because we used slightly different photometry parameters, we used the image from the archive with updated calibration, and because here we report a Vega-based magnitude, whereas the previous measurement was ST mag.  The second column in Table~\ref{tab:phot} lists the phase of the observations relative to the presumed time of explosion, which is just before the start of the 2012a event (this is 2012 Aug 4, or JD = 2456144, and is the same as $T_{\rm SN}$ in \citealt{smith14}).  This is the same as Day 0 in Figure~\ref{fig:lc}.

Figure~\ref{fig:lc} shows the late-time light curve of the point source at the position of SN~2009ip, including \hst\, photometry from Table~\ref{tab:phot}.  This is plotted as absolute magnitude, with an extinction correction applied (although the extinction correction is very small).  Following \citet{smith10}, we adopt a distance modulus of $m-M = 31.55$~mag and a foreground Galactic reddening of $E(B-V) = 0.019$~mag for NGC~7259.  It seems unlikely that there is a large amount of additional extinction and reddening local to the host galaxy, given the object's very remote location in the outskirts of NGC~7259 \citep{smith16sn09ip}. 

\begin{figure}
\begin{center}
\includegraphics[width=3.6in]{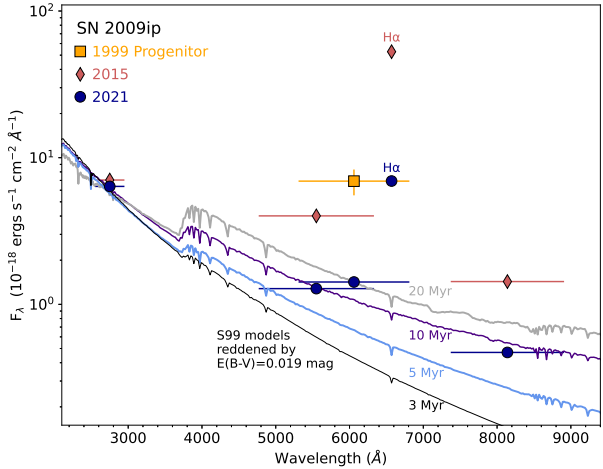}
\caption{The late-time SED of SN~2009ip in 2015 and 2021 compared to the 1999 progenitor and cluster models.  The F606W flux for the quiescent progenitor \citep{smith10} is shown with an orange square, whereas the late-time \hst \, photometry from 2015 and 2021 (see Table 1) is shown with red diamonds and blue circles, respectively.  The F657N flux is dominated by H$\alpha$ emission.  Simulated Starburst99 (S99) spectra for star clusters of 3, 5, 10, and 20~Myr are displayed for comparison (see text), after being reddened with $E(B-V) = 0.019$~mag, appropriate for the line-of-sight Milky Way extinction.}\label{fig:sed}
\end{center}
\end{figure}

Figure~\ref{fig:sed} displays the ultraviolet (UV)/optical SED of SN~2009ip in the F275W, F555W, F657N, and F814W filters at late times in 2015 (red) and 2021 (blue; also including F606W), as well as the 1999 progenitor in the F606W filter only (orange).  For comparison, Figure~\ref{fig:sed} also shows the simulated spectra of hypothetical underlying young star clusters with ages of 3, 5, 10, and 20~Myr made with Starburst99 models \citep{leitherer99}.  These simulated cluster spectra have been reddened slightly by $E(B-V) = 0.019$~mag and scaled to the observed F275W flux (i.e., the observed fluxes in Fig.~\ref{fig:sed} were not corrected for extinction, but the model spectra were reddened).  The SED is discussed in the next section. Figure~\ref{fig:sed} is very similar to Figure 2 from \citet{smith16sn09ip}, which showed similar Starburst99 models compared to the 2015 photometry.  The 2015 fluxes here are slightly different because we have measured the photometry in images with updated calibration, and using the same photometry parameters as for the new images obtained in 2021.

\section{RESULTS}\label{sec:results}

\subsection{Light curve}

Figure~\ref{fig:lc} shows the light curve of SN~2009ip, including late-time \hst\, photometry from Table~\ref{tab:phot} in all five filters observed at the recent epoch in December 2021:  F275W, F555W, F606W, F657N, and F814W.  The most notable result is that in the F606W filter, the late-time source at the position of SN~2009ip is now obviously fainter than the quiescent progenitor detected in F606W \hst/WFPC2 images in 1999.  The progenitor's absolute magnitude in 1999 is indicated by the horizontal blue bar, and at the latest 2021 epoch, SN~2009ip is now more than 1.2~mag fainter.  This is strong evidence that the luminous LBV-like progenitor did not survive the 2012 SN-like event.  Several aspects of the late-time behaviour in multiple filters are discussed below.

{\it Steady and smooth decline:}  Since falling from the peak of the 2012b event, SN~2009ip has only continued to fade and is now the faintest it has ever been in the optical.  Immediately after the 2012 event (around day 200), it was declining somewhat slower than the rate of $^{56}$Co decay for a $^{56}$Ni mass of 0.04~$M_{\odot}$ (shown by the dashed grey line in Fig.~\ref{fig:lc}).  During that time, its light curve and spectral properties were consistent with those of late-time SNe~IIn, and it had an underlying broad-line spectrum similar to that of SN~1987A \citep{graham14,smith14}.  Up to about day 1000 it continued to fade smoothly at a slower rate around $0.003$~mag d$^{-1}$, and spectroscopically it continued to resemble late-time interaction in SNe~IIn \citep{fox15,smith16sn09ip,graham17}.  Our new \hst\, photometry shows that from about day 1000 to day 3000, SN~2009ip has continued to fade smoothly and steadily in the optical, at an even slower rate. While the {\it HST} cadence is obviously sparse at these late times, the filters with more than two observations (F555W and F814W) show no significant deviation from a steady decline; there is no evidence of any rebrightening or irregularity in the fading rate.  From 2015 to 2021, the decline rates in the various optical filters are $0.00051 \pm 0.00001$~mag~d$^{-1}$ in F555W, $0.00068 \pm 0.00002$~mag~d$^{-1}$ in F606W, $0.00092 \pm 0.00001$~mag~d$^{-1}$ in F657N, and $0.00050 \pm 0.00001$~mag~d$^{-1}$ in F814W. (The UV is an exception, as discussed below.)  While this decline is much slower than radioactive decay, such slow decline rates are not at all unusual for SNe~IIn with late-time CSM interaction.  SNe~IIn span a wide diversity of late-time decay rates, ranging from some SNe~IIn that have essentially flat light curves for many years, like SN~2005ip \citep{smith09sn05ip,smith17,stritz12,fox20}, down to those that have only weak CSM interaction and faster decline rates that are difficult to distinguish from radioactive decay or light echoes.  Some SNe~IIn even rebrighten at late times, like SN~2006jd \citep{stritz12}. In any case, it seems as if SN~2009ip is leveling off and approaching a constant $V$ absolute magnitude of $-7.5$ or so.  In all three broad continuum filters, the object is now fainter than the progenitor, and the light curve has not exhibited any additional eruptive variability since the 2012b event.

{\it Constant optical colour:}   A very important property of the late-time fading is that all visual-wavelength filters seem to fade in-step, showing similar decline rates.   The F555W and F814W decline rates are identical within the uncertainties, indicating an unchanging $V-I$ colour as SN~2009ip fades.  The narrow F657N filter, which samples strong H$\alpha$ emission, shows a somewhat faster decline rate than the F555W and F814W continuum filters.  Note that the F606W filter bandpass includes H$\alpha$ emission as well, but diluted within the broad filter; this probably explains why the F606W decline rate is faster than F555W and F814W, but not as fast as the narrow F657N filter.
As we discuss later, the late-time ``continuum'' luminosity may actually be a combination of two sources: fading luminosity from the SN itself, plus a constant underlying continuum (perhaps from a host star cluster).  This would be consistent with the faster decline rates in filters that include H$\alpha$ emission, since the H$\alpha$ emission is dominated by CSM interaction in the fading SN, whereas the continuum may arise from both sources.  In any case, the constant $V-I$ colour is physically meaningful, because it shows that SN~2009ip is not getting redder as it fades below its progenitor's brightness.  The constant visual-wavelength continuum colour therefore indicates that {\it the fading source cannot be dimming because of increasing extinction from dust.}  This, in turn,  definitively rules out a scenario where the luminous progenitor has survived, but appears fainter now because it is obscured by dust that formed in the event (as is the case for classic LBVs like $\eta$~Car).   Whatever star was dominating the luminosity in 1999 is now gone, not hidden.

{\it UV vs.\ optical:}  While the visual-wavelength filters all fade together in-step, albeit with a somewhat faster decline rate in filters that include H$\alpha$, it is interesting that the source at the position of SN~2009ip barely fades at all in the UV.  The decline rate in the F275W filter from 2015 to 2021 is only $0.00005 \pm 0.00001$~mag~d$^{-1}$ (i.e., 10 times slower than F555W).  This means that whatever source dominates the UV flux is nearly constant, and cannot be the same as the source of the fading optical continuum and H$\alpha$ luminosity.  If the optical continuum were getting significantly bluer while the UV brightness stayed the same, then perhaps we could imagine a conspiracy wherein the temperature was increasing at just the right rate to keep the F275W magnitude unchanged while the optical source faded.  However, the constant $V-I$ colour rules this out.  The constant $V-I$ colour, combined with the relatively constant UV flux, therefore eliminates a hypothesis where a luminous star survived, but is fainter because it became hotter (for instance, if the 2012 event were the last puff of mass loss that propelled the surviving star into a hot post-LBV phase as a Wolf-Rayet star).  Instead, there must be two luminosity sources contributing to the lingering late-time flux, and so we next investigate the shape of the SED.

\subsection{SED: UV vs. Optical}

A striking result from the latest photometry is that all the optical filters are continuing to fade at a similar rate, whereas the UV flux in the F275W filter has barely changed at all since 2015.  This suggests that  {\it there are at least two different sources that contribute to the observed light}: a nonvariable source in the UV, and a fading optical source.  A straightforward conjecture would be that some light is emitted by the SN itself, which fades as late-time CSM interaction declines (especially since the optical continuum and H$\alpha$ seem to fade together), while some light is contributed by an unchanging host star cluster in which the SN progenitor was born.

It was already suggested that an underlying star cluster probably contributes to the late-time UV luminosity \citep{smith16sn09ip}, but constraints in the 2015 data were not very strict.  The fact that the optical flux has continued to fade while the UV hasn't now allows much tighter constraints on an underlying stellar population: the data indicate that the stellar population is quite blue and therefore young.  This is in qualitative agreement with previous claims that the progenitor was a very massive star \citep{smith10}, despite its remote location in the host galaxy \citep{smith16sn09ip}.

If we assume that the F275W flux observed in 2021 is dominated by the UV continuum of a young star cluster, then the F555W and F814W fluxes help to quantitatively constrain the age of that cluster.  Figure~\ref{fig:sed} shows the observed \hst\, fluxes in 2015 and 2021 compared to simulated spectra from the integrated light of young star clusters.  These correspond to clusters with ages of 3, 5, 10, and 20~Myr from the Starburst99 population-synthesis code \citep{leitherer99}, and they are scaled to be just below the most recent F275W flux in December 2021.  Since the F555W and F814W points are currently below the spectrum of a 10~Myr cluster (i.e., just under the purple line in Fig.~\ref{fig:sed}), it looks like any underlying star cluster must be younger than 10~Myr, implying an initial mass for the progenitor of SN~2009ip higher than 20~$M_{\odot}$.  Of course, this is an upper limit to the age and a lower limit to the initial mass.  This is because some SN light still contributes to the broad-band flux in F555W and F814W as well, including various emission lines (such as a pseudocontinuum of many Fe lines in F555W and the Ca~{\sc ii} near-IR triplet and other lines in F814W; see Fig.\ 3 of \citealt{smith16sn09ip}).  If there is any contribution of SN light to those broad filters, then the cluster light would need to be even fainter and bluer, pushing the underlying cluster to younger ages like 5~Myr or less, thus implying an initial mass of $\sim 50~M_{\odot}$ or more for the progenitor.  Also, we have only applied a reddening appropriate for the foreground line of sight through the Milky Way; if there is any additional extinction and reddening local to the host galaxy of SN~2009ip, then the cluster would need to be intrinsically even bluer and younger.

\begin{figure}
\begin{center}
\includegraphics[width=3.3in]{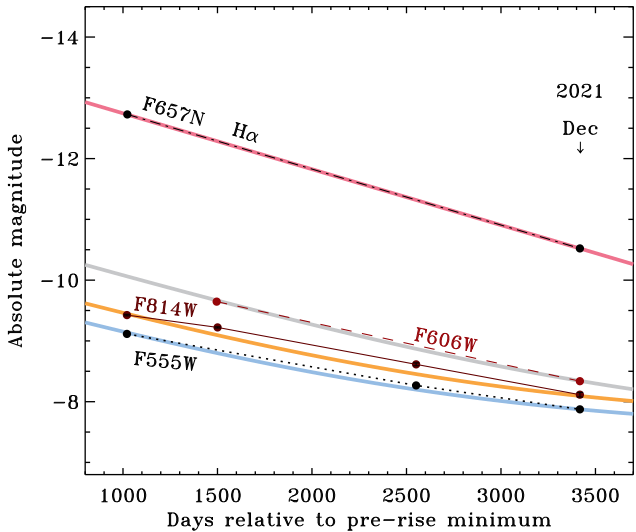}
\caption{Same as the light curve in Figure~\ref{fig:lc}, but zoomed-in on late times with \hst\, photometry.  A hypothetical decay rate is plotted for each filter.  For F657N (pink line), there are only two points, so this is a straight exponential decay with  $F \propto e^{t/1087}$ ($t$ in days).  The blue (F555W), orange (F814W), and grey (F606W) curves use the same exponential decay rate as F657N, but they have a constant flux added  to represent the unchanging light of a host cluster.  See text.}\label{fig:lc2}
\end{center}
\end{figure}

As time proceeds, we expect the CSM interaction luminosity of SN~2009ip to continue to fade, but if a cluster contributes to the light, then the optical source at the object's position should level off as the UV already seems to have done.  Figure~\ref{fig:lc2} explores this hypothesis.  This is the same as the light-curve plot in Figure~\ref{fig:lc}, except zoomed-in on only the late times where we have \hst\, photometry.  It displays four hypothetical fading curves for the four filters.   The pink line for F657N is an exponentially fading flux, with $F \propto e^{-t/1087}$ and $t$ in days, matching the decay rate of 0.00092~mag~d$^{-1}$ in this filter.  If we assume this to represent the true rate of fading of the CSM interaction luminosity, then we can predict how much cluster light needs to be added to the broad filters to match their decay rates.  The other three curves (blue for F555W, orange for F814W, and grey for F606W) adopt the same $F \propto e^{-t/1087}$ rate of fading for the SN~2009ip contribution to each filter, but here we add a constant flux level representing the unchanging contribution from an underlying star cluster, and then convert to magnitudes.  All three curves converge to this constant luminosity eventually (after another decade or so), which corresponds to absolute magnitudes of approximately  $-7.5$ in F555W and F606W, and $-7.65$ in F814W.  The fact that we can roughly match the observed decay rate of both the F555W and F606W filters using the same SN decay rate plus the same unchanging cluster continuum flux confirms our conjecture that the F606W filter traces the same continuum as F555W, but with H$\alpha$ emission adding to the flux.  These curves slightly underestimate the brightness of the intermediate points for F555W and F814W around days 1500 and 2550, but this is not too worrisome, since it is based on assuming a perfectly smooth decline connecting two observed F657N points. Without data in between, we don't know how F657N actually behaved.  There is no reason to expect a perfect exponential decline, since CSM interaction in decade-old SNe~IIn can be irregular owing to density variations in the CSM \citep{smith17,fox20}.  Obviously, the CSM interaction decay rate may change with time, in which case the final magnitudes in broad filters may differ somewhat from our simple estimate, but this demonstrates that a constant contribution from a cluster is plausible.

Thus, judging by the light curve, SN~2009ip may level off at an absolute $V$ magnitude of around $M_V \approx -7.5$ (or perhaps fainter if the cluster is younger than 5~Myr).   This would be much fainter than a super star cluster like R136 in 30~Dor (integrated $M_V \approx -9$~mag), but similar to the young massive cluster Tr~14 in the Carina Nebula ($M_V \approx -7.7$~mag).  The host cluster is likely older than Tr14 ($\sim 2$~Myr), however, since there is no longer a bright extended H~{\sc ii} region around SN~2009ip \citep{smith16sn09ip}.  This makes an age of 4--5~Myr seem likely for the progenitor. If so, the integrated light will asymptotically approach the level of the underlying star cluster, eventually providing definitive constraints on the age of the host stellar population among which the progenitor was born.  When it levels off, the inclusion of IR photometry and spectra may help to assess the properties of any red supergiants in the cluster, which may give important insight on the age and mass turnoff in the cluster \citep{beasor19}, as long as SN~2000ip did not form much dust.  UV spectra may reveal the earliest O-type stars that contribute to the cluster light.   We note that this may provide an interesting test of the progenitor's evolution.   For instance, if the underlying cluster has an age that points to an initial mass for SN~2009ip of $\sim 35$--40~$M_{\odot}$ (or if it contains red supergiants with initial masses below $\sim 30$~$M_{\odot}$), but the quiescent progenitor seemed to be a 50--80~$M_{\odot}$ star \citep{smith10}, then this might imply that the progenitor was made into a rejuvenated blue straggler by binary interaction, such as a merger or mass gainer.  This is  suggested based on its isolated environment \citep{smith16sn09ip}, and it would be consistent with implications from the environments of nearby LBVs \citep{st15,mojgan17}.

If or when the source does stop fading, it would be naive to interpret this leveling off as a surviving star.  It is already much fainter than the progenitor and blue --- and massive stars do not, after all, form in complete isolation.  Some massive evolved stars like LBVs may appear to be farther from O-type stars than expected for simplistic single-star scenarios \citep{st15}, but they are still found with relatively nearby OB associations within tens to hundreds of pc.  SN~2009ip has nothing else within a few kpc \citep{smith16sn09ip}.  In order for its progenitor to be truly isolated, one would need to assume that it was an extreme velocity runaway star to get it out to its current location.  But even with past binary evolution, this is a tall order given the apparent high mass and short lifetime of the progenitor.  Invoking a host star cluster is far more straightforward than invoking a bizarrely massive star that must have been a hypervelocity runaway to become so freakishly isolated.  Could the constant source be a surviving companion of SN~2009ip in a binary system?  Some companion candidates to other SNe have been revealed in late-time data \citep{fox14,fox22,ryder18,maund19,sun22}. If the companion is a main-sequence OB star, it would be too faint to contribute significantly; it would need to be a blue or yellow supergiant to produce the expected light level alone.  Having a companion that is also evolved is rare enough (because they would need to have nearly identical initial mass in order to have the same main-sequence lifetime), but that evolved supergiant would also need to be much fainter than its companion of the same mass.  This seems very unlikely.

Finally, we note that if a young star cluster contributes to the late-time luminosity, then this same cluster must also have contributed to the luminosity at the time the progenitor was detected in 1999.  This would, however, be a small correction.  A $\sim$5~Myr cluster shown in Figure~\ref{fig:sed}, for example, would contribute $< 10$\% of the flux in the F606W filter in 1999, implying a marginally lower luminosity for the progenitor of log($L$/L$_{\odot}$) $= 5.85$ instead of 5.9.  This would lower the inferred initial mass somewhat, from about 50--80~$M_{\odot}$ \citep{smith10} to about 45--70~$M_{\odot}$, depending on the bolometric correction.

\begin{figure}
\begin{center}
\includegraphics[width=3.1in]{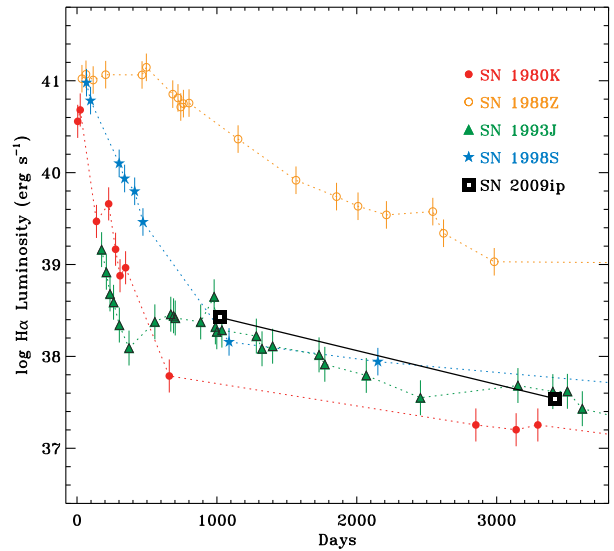}
\caption{The late-time decay in H$\alpha$ luminosity of SN~2009ip (heavy black squares and solid line) derived from F657N photometry, compared to H$\alpha$ luminosities from spectra for a few well-studied SNe with late-time CSM interaction.  This is adapted from Figure 6 of \citet{smith17}, including H$\alpha$ luminosities for SN~1980K (filled red circles), SN~1988Z (open orange circles), SN~1993J (green triangles), and SN~1998S (blue stars); see that paper for details.  The uncertainty for SN~2009ip is dominated by the adopted continuum flux, and the resulting error bar is smaller than the plotting symbol. }\label{fig:Ha}
\end{center}
\end{figure}

\subsection{SED: H{$\alpha$}}

At late times in both 2015 and 2021, the observed SED of SN~2009ip exhibits a very strong H$\alpha$ emission excess in the F657N filter, as compared to the broad F555W and F814W filters (we note that the broad F606W filter includes this H$\alpha$ emission, causing it to somewhat overestimate the underlying continuum level).   \citet{smith16sn09ip} already compared the 2015 SED to a contemporaneous spectrum of SN~2009ip, which did indeed reveal very strong, shock-broadened H$\alpha$ emission.  \citet{smith16sn09ip} also showed that the relative fluxes in F657N and the broad-band optical filters were consistent with expectations for a SN with strong late-time CSM interaction, where a forest of fainter emission lines produce a pseudocontinuum.  

The observed flux density in December 2021 through the narrow F657N filter is 6.9$\times$10$^{-18}$~erg~s$^{-1}$~cm$^{-2}$~\AA$^{-1}$.  One can infer a red continuum level by interpolating between the F555W and F814W fluxes (in fact, the 10~Myr cluster shown in Figure~\ref{fig:sed} gives a suitable guide for the red continuum shape), giving an expected 2015 continuum level near H$\alpha$ of $\sim$7.5$\times$10$^{-19}$~erg~s$^{-1}$~cm$^{-2}$~\AA$^{-1}$, yielding an excess that we may attribute to H$\alpha$ line emission of $\sim$6.2$\times$10$^{-18}$~erg~s$^{-1}$~cm$^{-2}$~\AA$^{-1}$.  The F657N filter has a width of 121~\AA\ (see the WFC3 Instrument Handbook\footnote{\tt https://hst-docs.stsci.edu/wfc3ihb/}), indicating a continuum-subtracted H$\alpha$ line flux of 7.5$\times$10$^{-16}$~erg~s$^{-1}$~cm$^{-2}$, an H$\alpha$ luminosity of 8800~L$_{\odot}\,(D/20.4~{\rm Mpc})^{2}$, or an H$\alpha$ emission equivalent width (EW) of $\sim$1000~\AA.  This EW is at the high end of values seen in the precursor eruptions of SN~2009ip, similar to those during its post-2012 decline when CSM interaction was very strong, and is typical of core-collapse SNe interacting with dense CSM at late times, as noted already by \citep{smith14}.  
Figure~\ref{fig:Ha} shows the late-time H$\alpha$ luminosity of SN~2009ip derived from our F657N photometry (with an assumed continuum level subtracted as described above), as compared to H$\alpha$ luminosities for a few well-studied SNe with observable CSM interaction in the decade after explosion \citep{smith17}.  Although less luminous than SN~1988Z (one of the strongest-interacting SNe~IIn), SN~2009ip is almost identical to the H$\alpha$ declines of SN~1998S and SN~1993J.  This comparison makes it very plausible that the late-time H$\alpha$ luminosity of SN~2009ip arises from CSM interaction.
On the other hand, the $\sim$1000 \AA \, H$\alpha$ EW of SN~2009ip is much stronger than typical H$\alpha$ EWs of quiescent LBVs (usually 100--200~\AA).   It exceeds the LBV with the strongest known H$\alpha$ EW, which is $\eta$~Car with an H$\alpha$ EW that is variable over the range 400--600~\AA\ \citep{smith03,stahl05}.

For interpreting the H$\alpha$ EW, a key difference between 2015 and 2021 is that if there is an underlying star cluster that dominates the UV flux in the F275W filter (as discussed above), then a larger fraction of the optical continuum must come from that cluster in 2021 (as compared to 2015).  This, in turn, indicates that the ``continuum'' flux contributed by the SN itself has faded even more than indicated by the light-curve decline rate, and therefore that 1000~\AA\ is a significant underestimate of the true H$\alpha$ EW in 2021.  (The EW could easily be 2000~\AA\ if the host cluster contributes half the continuum light, which it must for an age of 5~Myr).  This makes it highly unlikely that the broad H$\alpha$ seen at late times comes from the wind of a surviving LBV star, with a much more likely cause of the H$\alpha$ emission being late-time CSM interaction.  The H$\alpha$ EW in 2015 was about 2430~\AA.  It had risen steeply from only $\sim 100$~\AA\ at the luminosity peak in 2012 to an EW of more than 4000~\AA\ a year later in 2013 \citep{smith14}.  Thus, the H$\alpha$ EW weakened substantially from 2013 to 2015, and is now fading at a slower rate or leveling off.

\section{DISCUSSION AND CONCLUSIONS}\label{sec:conclusion}

We have presented and analysed \hst/WFC3 images of the source at the position of SN~2009ip taken almost a decade after the 2012a explosion, including photometry in UV and optical filters.  Here we list the main conclusions.

(1) At almost a decade after explosion, the source at the position of  SN~2009ip has now definitively faded below its quiescent progenitor seen 13~yr before explosion.  In particular, it is about 1.2~mag fainter in the same F606W filter as the progenitor detection, and is even fainter in F555W and F814W.  It is 6--7~mag fainter than its pre-SN eruptions.

(2)  For the past 7~yr, the source has steadily and continuously faded, showing no bumps in the light curve that might signify recurring eruptions.  Although the cadence has been admittedly sparse, all available data in multiple filters are consistent with nearly the same constant rate of decline.

(3)  All optical filters exhibit a similar steady decline, although filters including H$\alpha$ have a somewhat faster decline rate. Particularly noteworthy is that the F555W and F814W filters have the same decline rate of 0.0005~mag~d$^{-1}$ and have therefore maintained the same $V-I$ colour for the past 7~yr.  

(4) The constant $V-I$ colour as SN~2009ip fades is highly constraining.  It means that the source is not getting any redder as it fades, ruling out a hypothesis that a surviving star has become fainter than the progenitor because of dust formation or shifting to much cooler temperatures.  It also means that the source has not become bluer, eliminating the hypothesis that a surviving star has become fainter at visual wavelengths because it has become progressively hotter (as it would if a post-LBV were becoming a Wolf-Rayet star, for example).

(5)  The H$\alpha$ emission line has remained very strong, with a lower limit to the H$\alpha$ emission EW of $> 1000$~\AA.  This is $\ge$10 times stronger than the H$\alpha$ EW during the 2012b peak \citep{smith14}, making it unlikely that a light echo dominates the late-time source, indicating instead that late-time CSM interaction dominates.  If the observed continuum is a mix of SN light plus continuum from an underlying stellar population (see next points), then the true H$\alpha$ EW is even larger.  Such strong H$\alpha$ emission is not seen in LBV stars with luminous electron-scattering photospheres, but it is common in the late-time spectra of interacting core-collapse SNe.  The continuum-subtracted H$\alpha$ luminosity and decline rate of SN~2009ip in the decade after explosion are almost identical to those of SN~1998S and SN~1993J.

(6)  Although the visual-wavelength continuum shows no change in colour and all filters fade steadily at similar rates, the UV light sampled by the F275W filter exhibits barely any change at all, fading by only 0.11~mag in 7~yr.  Since the $V-I$ colour has not changed, this indicates that at least two distinct sources must contribute to the light now seen at the object's position: one is a constant source in the UV, and one is the fading SN.

(7)  The evolution of the late-time SED can be easily explained by the light of a fading SN~IIn combined with constant light from an underlying young star cluster.  Comparison of the current SED to Starburst99 models of the integrated spectrum constrains this cluster to be significantly younger than 10~Myr.  However, SN~2009ip is still fading.  As it continues to fade at visual wavelengths, this cluster will be constrained to progressively younger ages.  Judging by the current rate of fading, we expect that the source at the position of SN~2009ip may eventually settle down to an absolute $V$ magnitude of around $-7.5$, or perhaps even fainter if the cluster is younger than 5~Myr.  Such a young cluster would be consistent with the very massive progenitor star, and might be similar in mass to young Galactic clusters like Tr14 in the Carina Nebula (although somewhat older).  Multiwavelength photometry or spectroscopy in a few years when it levels off should be able to place tight constraints on the age of the stellar population in which the progenitor was born.  IR photometry to constrain any possible contribution from red supergiants may place the tightest limits on the age.

In summary, any source at the SN position has now become less luminous than the progenitor detected 13~yr before explosion.  Its visual-wavelength fading does not result because it is obscured by dust or has evolved to hotter temperatures.  The H$\alpha$ emission implies that ongoing CSM interaction still  contributes to the luminosity, while the SED suggests that the late-time light is actually a combination of two sources: (1) the fading CSM interaction from the dying transient, and (2) a constant blue source that dominates the unchanging UV emission, which is likely due to the integrated light of the young stellar population in which the SN~2009ip progenitor was born.

Taken at face value, this confirms expectations of a scenario involving a terminal core-collapse SN for SN~2009ip wherein the progenitor star is now dead. A hypothetical surviving star in a nonterminal scenario was expected to return to the progenitor's state, and this is now clearly ruled out. While the observations are fully consistent with the 2012 event of SN~2009ip being a terminal core-collapse SN, we must admit two caveats.  (1) expectations that a surviving source would ``return to normal'' after such an energetic event were perhaps overly simplistic and were not based on any physical model of nonterminal events. Nevertheless, if  one still wishes to invent nonterminal models for SN~2009ip that agree with the observations, one must now invoke additional ``epicycles" to explain the post-explosion faintness.   (2) While the fading presents compelling evidence that the 2012 explosion was a terminal event, a neutrino-driven explosion resulting from the collapse of an Fe core in a single massive star is not  necessarily a unique explanation for the event (even if it is a straightforward one).  The complex light curve of SN~2009ip, combined with the masking effects of CSM interaction, may motivate other interesting physical origins for a terminal SN-like event.  We consider some alternative interpretations below.

\subsection{Pulsational Pair Instability} 
Concerning nonterminal explosions, an obvious candidate mechanism is a pulsational pair instability (PPI) eruption \citep{woosley17}.  The PPI mechanism is frequently invoked by observers as a panacea for multipeaked SN light curves, for pre-SN eruptions and massive CSM shells, and for the hopes of a star lingering after an explosion. 
It has been invoked by several authors as a possibility for SN~2009ip \citep{pastorello13,fraser13,margutti14}.  However, while the PPI can produce diverse outcomes, it is perhaps not as malleable as some have suggested, and there are a few observed properties of SN~2009ip that contradict expectations for a PPI event (see \citealt{woosley17} and \citealt{renzo20} for detailed discussions of PPI eruption models over a range of initial masses).  For example, PPI models with $10^{50}$ erg eruptions generally predict a series of ejected shells with similar bulk expansion speeds of $\sim 2000$~km~s$^{-1}$ --- very different from the case of SN~2009ip, which had a series of weak and slow (500~km~s$^{-1}$) shell ejections followed by a single major pulse of 13,000~km~s$^{-1}$.  In fact, \citet{renzo20} find that over a wide range of masses in PPI models, the bulk ejection speeds of subsequent pulses are typically slower than the first pulse, opposite to the case of SN~2009ip.   

Also, in PPI models, the first pulse almost always ejects any remaining H envelope.  This is inconsistent with SN~2009ip, which had repeating mass-loss eruptions before 2012, but then had a major event in 2012 that exhibited strong and broad H$\alpha$ P~Cygni profiles from the H-rich ejecta of that event.  Clearly, the star still retained a considerable H envelope at the time of the 2012a/2012b explosion. Moreover, any surviving star after a PPI event would have no H left in its atmosphere, ruling out the idea that the late-time H$\alpha$ in SN~2009ip might be from a surviving LBV's wind emission. 

Depending on the star's initial mass and other parameters, the ensuing PPI eruption is not even accurately framed as a nonterminal event in many cases, since a collapse to a black hole often occurs during the main light-curve peak or very soon thereafter \citep{woosley17}. Models with a star that survives for decades or more are relatively rare and restricted to the highest initial masses.  The initial mass of 50--80~$M_{\odot}$ inferred for the progenitor of SN~2009ip \citep{smith10} overlaps only partly with the bottom of the initial mass range of 70--140~$M_{\odot}$ that is expected to experience the PPI \citep{woosley17}. However, models with time delays between pulses of a few weeks or more (needed for the multipeaked structure of SN~2009ip's 2012a/b event) require initial masses above 95~$M_{\odot}$ \citep{wh21}.   Finally, a PPI eruption also provides no explanation for disc-like asymmetry in the CSM.

\subsection{Stellar Merger}  
The strong asymmetry in the CSM of SN~2009ip indicated by polarization studies \citep{mauerhan14,reilly17} provides compelling evidence that binary interaction probably played an important role in the pre-SN mass loss.  The progenitor's very high luminosity \citep{smith10}, combined with its extremely remote environment \citep{smith16sn09ip}, might also suggest that the progenitor has evolved through binary interaction channels, making it a blue straggler.  The brief pre-SN luminosity spikes of SN~2009ip have already been compared to the periastron collisions seen before $\eta$~Car's 19th century eruption \citep{S11,smith11,sf11,sa14,sk13}, and binary mergers and common-envelope events are known to eject massive asymmetric CSM shells \citep{mp06,pejcha16,pejcha17,mp17,hillel17}.  Mergers have been invoked to explain several non-SN transients \citep{tylenda06,tylenda11,st06,nandez14,pejcha14,smith16n4490,bla20,bla21}, although these are all much fainter than SN~2009ip and have much slower expansion speeds.  Even a violent merger in an extremely massive star like $\eta$~Car \citep{smith18fast,smith18} was far fainter (only $M_V = -14$~mag) and less energetic ($\sim 10^{50}$~erg) than the 2012 event of SN~2009ip.  Also, all of these merger-induced transients become faint, red, and dust obscured as they decline from peak, unlike SN~2009ip.  A merger of two massive stars seems unable to explain SN~2009ip.  Instead, a larger reservoir of explosive energy seems to be required, somehow synchronized with this pre-explosion binary interaction.

\subsection{Pre-SN Binary Interaction}  
Here we highlight two plausible scenarios in which violent binary interaction events might produce eruptive asymmetric mass loss that precedes a much more energetic SN-like event.  In one case, a binary system that was previously not interacting may be triggered to begin periaston collisions in an eccentric binary, or may enter a common-envelope phase; \citet{sa14} proposed that this binary interaction may be synchronized with core collapse when the primary expands owing to increased instability in late nuclear burning phases.  The primary star may then die in a core-collapse SN and interact with the surrounding dense asymmetric CSM, producing a significantly polarized SN~IIn.  This scenario is consistent with all available data.  

In another scenario, a massive star may begin interacting with a compact object companion, triggering an energetic SN explosion when the two enter a common envelope and merge \citep{fw98,chevalier12}.    \citet{schroder20} have presented models for mergers between a massive star and a neutron star or black hole, arguing that they can indeed produce terminal high-energy explosions powered by CSM interaction that resemble SNe~IIn.  In that study, mergers with a black hole seem to be favoured for producing more-luminous events like SN~2009ip.  In either scenario, the resulting display may not conform to traditional expectations for normal core-collapse SNe from lower-mass stars.  Both these unusual merger events and regular SNe result from the deposition of a large amount of energy inside a massive star's envelope, and they may be difficult to distinguish observationally when CSM interaction is involved, as the debate over SN~2009ip illustrates.

If not true ``supernovae,'' what shall we call these SNe~IIn that may result from terminal explosive mergers, and what are their distinguishing observables?  This is perhaps a question of branding rather than understanding the underlying physics, and there may be no shortage of new names to propose in this field of inquiry.   Regardless of what name we ultimately give to it, the 2012 event of SN~2009ip appears to have been a $\sim$10$^{51}$~erg terminal explosion that obliterated its massive previously erupting progenitor, which we suggest exploded while in the midst of violent binary interaction.

\section*{Acknowledgements}

Support was provided by the National Aeronautics and Space Administration (NASA) through {\it HST} grants GO-13787, AR-14295, GO-15166, and GO-16649 from the Space Telescope Science Institute, which is operated by AURA, Inc., under NASA contract NAS5-26555.  N.S.\ thanks S.\ Woosley for interesting discussions.
A.V.F. is grateful for support from the U.C. Berkeley Miller Institute for Basic Research in Science (where he was a Miller Senior Fellow) and the Christopher R. Redlich Fund.

\section*{Data Availability}

The data underlying this article will be shared on reasonable request
to the corresponding author.  {\it HST} data will be nonproprietary and available from the public archive.

\scriptsize
\bibliographystyle{mnras}
\bibliography{ref}
\label{lastpage}
\end{document}